\documentclass[12pt]{iopart}
\begin{document}
\article[]{}{Relations between parameters of the Hamiltonian and Neel-type states in the anisotropic Heisenberg model}
\author{Pavel Babaian and Gennady Koval}
\address{Faculty of Physics, Lomonosov Moscow State University, Moscow, 119991 Russia}
\vspace{10pt}
\begin{abstract}
This article investigates Neel-type states in the anisotropic Heisenberg model with an external field. For arbitrary spin $s$ and arbitrary dimension of the space $d$ we find the expressions relating the angles which define the directions of spin polarisation to the parameters of the Hamiltonian.
\end{abstract}
\begin{indented}
\item[Keywords: $XYZ$ model, factorized states, quantum spins, $d$-dimensional lattice.
]
\end{indented}
\section{Introduction}
The study of quantum spin systems is of considerable interest due to their relevance in fields like quantum information theory \cite{7b}. Such systems are often described by means of anisotropic Heisenberg model. In the general case, the eigenstates of this model cannot be analytically found, so the factorised states, which are eigenstates only in a limited parameter region, are of special interest.\\
\qquad The problem of factorised eigenstates was first investigated for a one-dimensional spin chain \cite{1b}. Such states play a role in quantum information protocols where system properties such as entanglement are important \cite{4b,8b, 10b, 11b}. Factorised states are primarily of interest because they are unentangled.  This interest, which has arisen relatively recently, has led to the papers investigating factorised states in models with higher spin values \cite{2b,3b}, with general two-spin interactions of arbitrary range \cite{9b}, and in models with higher space dimensions \cite{12b}.\\ 
\qquad  It is worth to note separately the work \cite{4b}, which addresses the issue of engineering such states on experiment. Therefore, from a practical point of view, it may be of importance to know the relation between the parameters of the Hamiltonian and the orientation of spins in the lattice sites. Such relations have been obtained only for some special cases \cite{1b, 13b}.\\
\qquad  In this work we study the spin-$s$ $XYZ$ model in an external arbitrary field on a $d$-dimensional cubic lattice with homogeneous nearest neighbours interaction. We obtain the relation between the parameters of the Hamiltonian and the angles characterising the Neel states.
\section{Model description}
Let's consider a $d$-dimensional cubic lattice of size $L_1\times L_2\times...\times L_d$, where all numbers $L_i$ are even. The Hamiltonian of spin-$s$ $XYZ$ model in an external field of arbitrary direction on such lattice reads
\begin{equation}\label{1}
    \mathcal{H}=\sum\limits_{\langle\mathbf{l},\mathbf{m}\rangle}(J_xS_\mathbf{l}^xS_\mathbf{m}^x + J_yS_\mathbf{l}^yS_\mathbf{m}^y + J_zS_\mathbf{l}^zS_\mathbf{m}^z)-\sum\limits_{\mathbf{l}}\mathbf{h}\mathbf{S}_\mathbf{l} , 
\end{equation}
where \textbf{l} and \textbf{m} are vector $d$-dimensional indices with integer components, which indicate a site on the lattice. In the exchange part of the Hamiltonian $\langle\mathbf{l},\mathbf{m}\rangle$ runs over the pairs of the nearest neighbours. $S^x_\mathbf{l}, S^y_\mathbf{l}, S^z_\mathbf{l}$ -- spin operators. $J_x, J_y, J_z$ are exchange integrals characterising for interspin interaction and $\textbf{h}$ is an external field.\\
\qquad In this model, if condition
\begin{equation}\label{2}
\fl    \qquad\qquad\frac{h^2_x}{(J_x+J_y)(J_x+J_z)}+\frac{h^2_y}{(J_y+J_x)(J_y+J_z)}+\frac{h^2_z}{(J_z+J_x)(J_z+J_y)} = (2ds)^2
\end{equation} is satisfied, the Hamiltonian \eref{1} has two Neel-type eigenvectors with the corresponding energy per site $\epsilon=-ds^2(J_x+J_y+J_z)$.\\
\qquad This result was first derived for a one-dimensional spin chain \cite{1b}. In \cite{12b} this result was obtained when considering the case $d=2$ and $s=\frac{1}{2}$.\\
\qquad  In general case of arbitrary spin $s$ and arbitrary dimension of the space $d$ the condition \eref{2}, under which Hamiltonian \eref{1} has Neel-type eigenvectors, is a consequence of the fact that \eref{1} can be represented as a sum of two-spin Hamiltonians
\begin{equation}\label{3}
    \qquad\qquad\qquad\qquad\mathcal{H}=\sum\limits_{\mathbf{l}}\sum\limits_{i=1}^d\mathcal{H}_{\mathbf{l},\mathbf{l}+\mathbf{e}_{i}},
\end{equation}
where 
\begin{equation}\label{4}
\fl {\mathcal{H}_{\mathbf{l},\mathbf{m}} =J_xS_{\mathbf{l}}^xS_{\mathbf{m}}^x + J_yS_{\mathbf{l}}^yS_{\mathbf{m}}^y + J_zS_{\mathbf{l}}^zS_{\mathbf{m}}^z-\frac{1}{2d}[h_x(S^x_{\mathbf{l}}+S^x_{\mathbf{m}})+h_y(S^y_{\mathbf{l}}+S^y_{\mathbf{m}})+h_z(S^z_{\mathbf{l}}+S^z_{\mathbf{m}})]}
\end{equation}
and $\mathbf{e}_i$ - basis vector of the lattice along the direction $i$. \\
\qquad  In the next section from consideration of the Hamiltonian \eref{4} we find expressions which relate Neel state's parameters to parameters of the Hamiltonian in general case.
\section{Angles and parameters}
\subsection{Spin-$\frac{1}{2}$}
In the case $s=\frac{1}{2}$ the matrix of the Hamiltonian \eref{4} has a size $4\times4$, which allows us to find all its eigenvalues explicitly. In arbitrary space dimension $d$ and spin $s=\frac{1}{2}$, Neel states have the form\\
$|N\rangle=\prod\limits_{\mathbf{m}}|\psi_\mathbf{m}\rangle$,\\
$|\psi_\mathbf{m}\rangle=a_\mathbf{m}|\uparrow\rangle + b_\mathbf{m}|\downarrow\rangle$.\\
The coefficients $a_\mathbf{m}$ and $b_\mathbf{m}$  express as follows\\
$a_\mathbf{m}=\cos(\frac{\theta_\mathbf{m}}{2})e^{-i\frac{\phi_\mathbf{m}}{2}}$, $b_\mathbf{m} = \sin(\frac{\theta_\mathbf{m}}{2})e^{i\frac{\phi_\mathbf{m}}{2}},$\\
where $\theta_\mathbf{m}$ and $\phi_\mathbf{m}$ are the polar and azimuthal angles, respectively, which satisfy the conditions
\begin{eqnarray}
   \qquad\quad \theta_\mathbf{m}=\frac{1+(-1)^{\sum\limits_{j=1}^dm_j}}{2}\theta_1+\frac{1-(-1)^{\sum\limits_{j=1}^dm_j}}{2}\theta_2, \label{20} \\
    \qquad\quad\phi_\mathbf{m}=\frac{1+(-1)^{\sum\limits_{j=1}^dm_j}}{2}\phi_1+\frac{1-(-1)^{\sum\limits_{j=1}^dm_j}}{2}\phi_2.\label{21}
\end{eqnarray}
This conditions are equivalent to the fact that two sublattices can be distinguished on the whole lattice on which the spins point in directions defined by angles $(\theta_1,\phi_1)$ on one sublattice and $(\theta_2,\phi_2)$ on the other.\\
$\textbf{\qquad}$We write the eigenvalue equation of the Hamiltonian \eref{4} for the Neel states
\begin{equation}\label{5}
    \qquad\qquad\mathcal{H}_{\mathbf{l},\mathbf{l}+\mathbf{e}_{i}}|\psi_\mathbf{l}\rangle|\psi_{\mathbf{l}+\mathbf{e}_i}\rangle=\frac{\epsilon}{d}|\psi_\mathbf{l}\rangle|\psi_{\mathbf{l}+\mathbf{e}_i}\rangle,
\end{equation}
where $\epsilon$ is the energy per site and the coefficient $\frac{1}{d}$ arises due to the presence of summation at index $i$ in \eref{3}. It follows from the form of the Hamiltonian $\mathcal{H}_{\mathbf{l},\mathbf{l}+\mathbf{e}_{i}}$ and conditions \eref{20} and \eref{21} that we can simplify the appearance of the formulas by the following substitution of notations $\mathcal{H}_{\mathbf{l},\mathbf{l}+\mathbf{e}_{i}}\rightarrow \mathcal{H}_{1,2}$, $|\psi_\mathbf{l}\rangle \rightarrow|\psi_1\rangle$ and $|\psi_{\mathbf{l}+\mathbf{e}_i}\rangle \rightarrow|\psi_2\rangle$. Substitution $|\psi_{1,2}\rangle$ into equation \eref{5} leads to the following system
\begin{equation}\label{6}
\cases{
   \frac{J_z}{4}-\frac{h_z}{2d} +\frac{J_x-J_y}{4}z_1z_2-\frac{h_x-ih_y}{4d}(z_1+z_2)=\frac{\epsilon}{d},
   \\
   \frac{J_x-J_y}{4}\frac{1}{z_1z_2}+\frac{J_z}{4}+\frac{h_z}{2d}-\frac{h_x+ih_y}{4d}(\frac{1}{z_1}+\frac{1}{z_2})=\frac{\epsilon}{d}, 
   \\
   -\frac{h_x+ih_y}{4d}\frac{1}{z_2}-\frac{h_x-ih_y}{4d}z_1-\frac{J_z}{4}+\frac{J_x+J_y}{4}\frac{z_1}{z_2}=\frac{\epsilon}{d}, 
   \\
   -\frac{h_x+ih_y}{4d}\frac{1}{z_1}-\frac{h_x-ih_y}{4d }z_2-\frac{J_z}{4}+\frac{J_x+J_y}{4}\frac{z_2}{z_1}=\frac{\epsilon}{d},}
\end{equation}
where $z_1=\tan(\frac{\theta_1}{2})e^{i\phi_1}$ and $z_2=\tan(\frac{\theta_2}{2})e^{i\phi_2}$. It follows from this system that for the spin $s=\frac{1}{2}$ if condition \eref{2} is satisfied, the Neel states are eigenstates of the Hamiltonian (4) with the corresponding energy per site $\epsilon=-d\frac{(J_x+J_y+J_z)}{4}$. \\
\qquad  To investigate the relations between angles and parameters, we substitute $\epsilon=-d\frac{(J_x+J_y+J_z)}{4}$ into the system \eref{6}. The substitution reduces the number of equations to three, since two equations coincide, and the system \eref{6} transforms into
\begin{equation}\label{7}
 \cases{
   (J_x+J_y+2J_z-2\frac{h_z}{d}) +(J_x-J_y)z_1z_2-\frac{h_x-ih_y}{d}(z_1+z_2)=0,
   \\
   J_x-J_y+(J_x+J_y+2J_z+2\frac{h_z}{d})z_1z_2-\frac{h_x+ih_y}{d}(z_1+z_2)=0, 
   \\
   -\frac{h_x+ih_y}{d}-\frac{h_x-ih_y}{d}z_1z_2+(J_x+J_y)(z_1+z_2)=0,}
\end{equation}
\subsubsection{Angles in terms of parameters\\}
Consider the system \eref{7} as a system of equations for variables $z_1$ and $z_2$. From this system \eref{7} follows the condition \eref{2}, which ensures the consistency of the equations. Also from \eref{7} we obtain expressions for the sum and product of $z_1$ and $z_2$
\begin{equation}\label{8}        \fl\qquad\qquad\cases{z_1z_2=\frac{h_x^2+h_y^2-d^2(J_x+J_y)(J_x+J_y+2J_z-\frac{2}{d}h_z)}{d^2(J_x^2-J_y^2)-(h_x-ih_y)^2}\equiv\alpha\\
z_1+z_2=2d\frac{(J_x-J_y)(h_x+ih_y)-(h_x-ih_y)(J_x+J_y+2J_z-2h_z)}{d^2(J_x^2-J_y^2)-(h_x-ih_y)^2}\equiv\beta}
\end{equation}
\qquad Then expressions for $z_1$ and $z_2$ take the form 
\begin{equation}\label{9}
    \qquad\qquad\qquad\qquad z_{1,2}=\frac{\beta\pm\sqrt{\beta^2-4\alpha}}{2}
\end{equation}
\qquad  Modulus and argument of $z_{1,2}$ results in the expressions for the angles $\theta_{1,2}$ and $\phi_{1,2}$, obtaining which was one of our aims.
\subsubsection{ Parameters in terms of angles\\}
 The system \eref{7} can also be considered as a system with respect to the parameters of the Hamiltonian at given angles. We consider separately imaginary and real parts from each of the equations, which results in the homogeneous system of 6 linear equations with respect to 6 parameters
\begin{equation}\label{10}
 \cases{
   J_x+J_y+2J_z-\frac{2}{d}h_z+(J_x-J_y)\gamma-h_x\frac{\delta}{d}-h_y\frac{\zeta}{d}=0,
   \\
   (J_x-J_y)\chi-h_x\frac{\zeta}{d}+h_y\frac{\delta}{d}=0,\\
   J_x-J_y+(J_x+J_y+2J_z+\frac{2}{d}h_z)\gamma-h_x\frac{\delta}{d}+h_y\frac{\zeta}{d}=0, 
   \\
   (J_x+J_y+2J_z+\frac{2}{d}h_z)\chi-h_x\frac{\zeta}{d}-h_y\frac{\delta}{d}=0,\\
   -h_x\frac{(1+\gamma)}{d}-h_y\frac{\chi}{d}+(J_x+J_y)\delta=0,\\
   -h_x\frac{\chi}{d}-h_y\frac{(1-\gamma)}{d}+(J_x+J_y)\zeta=0,} 
\end{equation}
where notation are\\
$\gamma=\tan(\frac{\theta_1}{2})\tan(\frac{\theta_2}{2})\cos(\phi_1+\phi_2)$,\\
$\delta=\tan(\frac{\theta_1}{2})\cos(\phi_1)+\tan(\frac{\theta_2}{2})\cos(\phi_2)$,\\
$\chi=\tan(\frac{\theta_1}{2})\tan(\frac{\theta_2}{2})\sin(\phi_1+\phi_2)$,\\
$\zeta=\tan(\frac{\theta_1}{2})\sin(\phi_1)+\tan(\frac{\theta_2}{2})\sin(\phi_2)$.\\
\qquad  Note that the system \eref{10} is consistent since its determinant is 0 automatically without any additional conditions. Since the rank of the matrix of system \eref{10} is 5, the solution space of such a system is one-dimensional.\\
\qquad  The solution of the system \eref{10} is determined to a constant multiplier. Let us choose $J_x$ as a fixed parameter, then the solution of \eref{10} takes the form 
\begin{equation}\label{11}
\qquad\qquad\left(
\begin{array}{c}
J_x \\
J_y \\
J_z \\
h_x \\
h_y \\
h_z
\end{array}
\right)
=J_x
\left(
\begin{array}{c}
1\\
        \frac{\chi(\chi^2+\gamma^2+\delta^2-\zeta^2-1)-2\gamma\delta\zeta}{\chi(\chi^2+\gamma^2-\delta^2+\zeta^2-1)+2\gamma\delta\zeta}\\
        -\frac{\chi(\chi^2+\gamma^2+\gamma(\zeta^2-\delta^2)-\delta\chi\zeta-1)+(1+\gamma^2)\delta\zeta}{\chi(\chi^2+\gamma^2-\delta^2+\zeta^2-1)+2\gamma\delta\zeta}\\
        \frac{2\chi(\delta(\gamma-1)+\chi\zeta)}{\chi(\chi^2+\gamma^2-\delta^2+\zeta^2-1)+2\gamma\delta\zeta}d\\
        -\frac{2\chi(\zeta(1+\gamma)-\delta\chi)}{\chi(\chi^2+\gamma^2-\delta^2+\zeta^2-1)+2\gamma\delta\zeta}d\\
        \frac{\delta^2\chi(1-\gamma)+\delta\zeta(\gamma^2-\chi^2-1)+\chi\zeta^2(1+\gamma)}{\chi(\chi^2+\gamma^2-\delta^2+\zeta^2-1)+2\gamma\delta\zeta}d
\end{array}
\right)
\end{equation}
\qquad The expressions for the parameters \eref{11} satisfy condition \eref{2} and there are no other solutions.
\subsection{Arbitrary spin}
We generalise the results obtained for spin $s=\frac{1}{2}$ to arbitrary values of spin. For this purpose we will use the theorem formulated in \cite{1b}. The formulation of the theorem is as follows: If $\mathcal{H}$ has a Neel eigenstate for
exchange constants $\textbf{J}$, and field $\textbf{h}$ for $s=\frac{1}{2}$, then $\mathcal{H}$ for arbitrary $s$ has also a Neel eigenstate for the same exchange constants and field $2s\textbf{h}$. It follows from the theorem that the condition for the existence of Neel states in the case of arbitrary spin $s$ is condition \eref{2}. Although the theorem was formulated in \cite{1b} for the one-dimensional case, its proof is directly generalised to the case of arbitrary dimension $d$.\\
\qquad Then the result \eref{8} is modified as follows
\begin{equation}\label{12}
\fl\qquad\qquad    \cases{
        z_1z_2=\frac{h_x^2+h_y^2-(2sd)^2(J_x+J_y)(J_x+J_y+2J_z-\frac{1}{ds}h_z)}{(2ds)^2(J_x^2-J_y^2)-(h_x-ih_y)^2}\equiv\alpha_s,\\
        z_1+z_2=4ds\frac{(h_x+ih_y)(J_x-J_y)-(h_x-ih_y)(J_x+J_y+2J_z-\frac{1}{ds}h_z) }{(2ds)^2(J_x^2-J_y^2)-(h_x-ih_y)^2}\equiv\beta_s.}
\end{equation}
\qquad Accordingly, the expression of the parameters of the Neel states in terms of the Hamiltonian parameters takes the form 
\begin{equation}\label{13}
   \qquad\qquad\qquad\qquad z_{1,2}=\frac{\beta_s\pm\sqrt{\beta_s^2-4\alpha_s}}{2}.
\end{equation}
\qquad The relation of the parameters of the Hamiltonian in terms the parameters of the Neel states transforms into
\begin{equation}\label{14}
\qquad\qquad\left(
\begin{array}{c}
J_x \\
J_y \\
J_z \\
h_x \\
h_y \\
h_z
\end{array}
\right)
=J_x
\left(
\begin{array}{c}
1\\
        \frac{\chi(\chi^2+\gamma^2+\delta^2-\zeta^2-1)-2\gamma\delta\zeta}{\chi(\chi^2+\gamma^2-\delta^2+\zeta^2-1)+2\gamma\delta\zeta}\\
        -\frac{\chi(\chi^2+\gamma^2+\gamma(\zeta^2-\delta^2)-\delta\chi\zeta-1)+(1+\gamma^2)\delta\zeta}{\chi(\chi^2+\gamma^2-\delta^2+\zeta^2-1)+2\gamma\delta\zeta}\\
        \frac{4\chi(\delta(\gamma-1)+\chi\zeta)ds}{\chi(\chi^2+\gamma^2-\delta^2+\zeta^2-1)+2\gamma\delta\zeta}\\
        -\frac{4\chi(\zeta(1+\gamma)-\delta\chi)ds}{\chi(\chi^2+\gamma^2-\delta^2+\zeta^2-1)+2\gamma\delta\zeta}\\
        \frac{(\delta^2\chi(1-\gamma)+\delta\zeta(\gamma^2-\chi^2-1)+\chi\zeta^2(1+\gamma))2ds}{\chi(\chi^2+\gamma^2-\delta^2+\zeta^2-1)+2\gamma\delta\zeta}
\end{array}
\right)
\end{equation}
\qquad $\gamma$, $\delta$, $\chi$ and $\zeta$ are the same ones that were introduced for the spin $s=\frac{1}{2}$. Also the expression \eref{14} still satisfies condition \eref{2}.
\section{Conclusion}
For arbitrary spin $s$ and arbitrary dimension of the space $d$ we have obtained  the relations between parameters of the Hamiltonian and angles characterising Neel states. As an intermediate result, a condition for the existence of Neel states was obtained for the case of arbitrary dimension of the space, which is a generalisation of this condition from \cite{1b}.\\
\qquad These results can possibly be used for modelling Hamiltonians with a given spin polarisation. In this work we have not addressed the problem of whether Neel states are ground states. This problem has been investigated partially in the cases of a one-dimensional spin chain with antiferromagnetic ordering \cite{1b} and a spin-$\frac{1}{2}$ lattice of arbitrary dimension with a one-dimensional external field \cite{8b}. We are currently working on finding out broader conditions on the parameters than in \cite{1b, 8b}, whose fulfilment guarantees that Neel states are ground states.

\end{document}